\documentstyle[12pt]{article}
	\makeatletter
	\@addtoreset{equation}{section}
	\makeatother
	
\begin{document}	
         
	\newcommand\beq{\begin{equation}}
	\newcommand\eeq{\end{equation}}
	\newcommand\en{energy momentum tensor}
	\newcommand\beqa{\begin{eqnarray}}
	\newcommand\eeqa{\end{eqnarray}}
	
	\tolerance=10000
	\vskip 1truecm
	\begin{center}
	{\LARGE Consistent two--dimensional chiral gravity}\\
	\vskip 2truecm
	{\large A.Smailagic\footnote{E-mail address: 
	anais@ictp.trieste.infn.it}},\\
	\medskip 
	{\it International Center for Theoretical Physics,\\
	 Miramare, 34014-Trieste, Italy,}\\
	and,\\
	{\it Dipartimento di Fisica Teorica dell'Universit\`a,\\
         Strada Costiera 11, 34014-Trieste, Italy,}\\ 
	\vskip 1truecm
	{\large E.Spallucci\footnote{E-mail address:
	spallucci@vstst0.ts.infn.it}}\\
	\medskip
	{\it Dipartimento di Fisica Teorica dell'Universit\`a,\\
	 Strada Costiera 11, 34014-Trieste, Italy,\\
	 Istituto Nazionale di Fisica Nucleare, Sezione di Trieste,\\
         Strada Costiera 11, 34014-Trieste, Italy,}

	\thispagestyle{empty}
	\vfill
	\begin{abstract}

	We study chiral induced gravity in the light--cone gauge and show that 
	the theory is consistent for a particular choice of chiralities.
	The corresponding Kac--Moody central charge has no forbidden
	region of complex values. 
	Generalized analysis of the critical exponents is given and
	their relation to the  $SL(2,R)$ vacuum states is elucidated.
	All the parameters containing information about the theory
	can be traced back to the characteristics of the group of residual
	symmetry in the light--cone gauge. 
	
	\end{abstract}
        \end{center}
\newpage	

\section{Introduction}

Two-dimensional induced gravity has become an interesting field of 
investigations since the early
days of Polyakov functional integral formulation of string theory \cite{Pol1}. 
Such a description enables a thorough investigation of the string dynamics 
only at the critical dimension, where 
there is no need to worry about gravitational quantum
effects. On the other hand, the occurrence of the anomaly in 
off--critical string
models makes unavoidable the presence of gravitational quantum effects,  
 and introduces a further technical problem in actual calculations. 
A ``proper'' gauge choice becomes an essential step in the formal
developments of computations.
Conventionally, off--critical models have been dealt with   in the 
conformal gauge \cite{gn}. 
The advantage of choosing such a gauge 
is to give the effective action a local form, i.e. the effective
dynamics of the Weyl degree of freedom is described by 
Liouville local field theory.  
However, there is a double price to pay: first, quantization becomes 
non-trivial due to the dependence on the Weyl factor; second, one
is faced with the, so far unresolved, problem \cite{tsey}
 to introduce, at the perturbative level, a
suitable regularization procedure.  
Therefore, the choice of the conformal gauge exchanges non--locality
with all the quantization problems of the Liouville field theory \cite{curt}.
Recently, some progress has been made
in the quantization of Liouville theory, as well as in treating the
strong coupling regime in the conformal gauge \cite{ger95},\cite{otto}.
Under this respect, 
it is somehow amazing that the light--cone gauge choice  is so
``clever'' to circumvent these difficulties \cite{Pol3} !
The regularization problem does not exist. The effective action, though
apparently in non--local form, can be handled and the model is
completely solvable. Furthermore, an unexpected residual $SL(2,R)$ symmetry
shows up. Due to the solvability of the model, the Kac--Moody central charge
can be renormalized in closed form and various {\it critical exponents}  
can be evaluated. Then, it is possible to compare them with the corresponding
quantities in statistical physics \cite{Pol4}, 
and verify a complete agreement among
certain non--critical string theories and some definite statistical models. 
However, the allowed  range of values for the central charge suffers from
a gap corresponding to complex, unphysical, values and forbids the
extrapolation of low dimensional results, corresponding 
to various statistical models, to higher dimensional ones 
relevant to string models.
This gap can be narrowed down by exploiting $N=1$ supersymmetry  and even
avoided in the $N=2$ supersymmetric model \cite{Pol5}. 

The motivation of the present paper is drawn from the known fact that
 chiral induced gauge theories \cite{jr}, though
anomalous, can be consistent under certain conditions. With the same hope,
we shall investigate the light--cone gauge structure of the chiral
induced gravity. This choice is motivated by the presence 
of a Lorentz anomaly which cannot be removed
by any regularization procedure \cite{anais1}. As a result, the Lorentz mode
shows up in the spectrum as a physical degree of freedom associated with 
an arbitrary 
parameter interpolating over different regularization schemes.\\
 One of the objectives of this paper is to employ the
freedom introduced by the regularization parameter with the hope
to eliminate the region of complex values of the central charge, 
as  an alternative way to the use of $N=2$ supersymmetry in non chiral models.\\
Another goal is to investigate critical exponents within light--cone
gauge description, and connect them to the characteristics of the residual 
symmetry group.\\
Finally, we would like to establish complete correspondence among
the parameters characterizing induced gravity models in the
light--cone and conformal gauges. While carrying on
this program, we have found  intriguing possibility of a  generalization
of various formulae, that could lead to a simple description of  
all the known chiral and non--chiral gravity models,
as well as their supersymmetric versions. Instead of bothering with a complex 
and different structures of various models,
and other technical details, we hope to absorb all the essential
characteristics of the mentioned models within {\it a couple of general 
parameters} from which one can work out the details of any
desired model. This, certainly, would be an interesting simplification
in handling the large number of specific results throughout the
literature. \\
The paper is planned as follows.\\
In Sect.2 we discuss the $N=0$ chiral gravity  model, and its classical
symmetries {\it before} any gauge choice. The purpose of this section is to 
offer some technical details and stress the differences between chiral 
and non chiral models. Since we are using a symmetric version
of the induced action we recall the connection to the asymmetric form of
the same quantity,
more often appearing in the literature. This model is more involved
than the non chiral one, due to the explicit presence of the spin connection.
We describe conservation laws and the corresponding symmetries assuming
the zweibein as the basic variable in the model. 
Our approach leads to an easy comparison of
the components of the energy--momentum tensor in different gauges. \\
In Sect.3 we give a detailed description of physical quantities in
the light--cone gauge, and discuss the residual
symmetries. We show that some of the components of the energy--momentum
tensor are generators of these residual symmetries, and can be described
in the Sugawara form. The remaining components produce equations of motion
for the dynamical components of the zweibein. Then, we study the general 
properties of the roots of the quadratic equations that describe the 
renormalized $SL(2,R)$ Kac--Moody central charge. 
    We also analyze the definition
of the renormalized gravitational scaling dimension and the string
susceptibility coefficient in terms of the scaling weights of the $SL(2,R)$
currents. In this way, we obtain a very simple meaning of these quantities
and the relations to scale operators connecting various physical states. We
also obtain a generalized form of the KPZ equation.\\
In Sect.4 we  investigate the connection between the characteristic parameters
of the conformal and light--cone gauge and establish a number of interesting 
general relations.\\
In Sect.5 we analyze various restrictions on the number of matter fields
and a free parameter characterizing the chiral model, resulting from
the requirement of reality of the parameters characterizing conformal
and light--cone gauge. At this point essential difference between chiral
and non--chiral model is pointed and discussed.\\ 
Finally, Sect.5 is devoted to a discussion of our results. 
We list a number of improvements and new results in chiral models.

\section{General description of the chiral model}

Supersymmetric and non--supersymmetric versions
of the non--chiral models have been studied extensively and details can be 
found in the literature \cite{Pol3},\cite{Pol5},\cite{distl}. Details of 
chiral models are less known and we shall give a comprehensive description
of the structure of their  non--supersymmetric version. In this way we avoid
complications due to the superspace description \cite{cqg}, while maintaining 
all the important  characteristics necessary for our discussion.  

The model is described by the induced action  resulting from the
integration of matter fields \cite{anais2}

\begin{eqnarray}
S_{symm.}&=&
{1\over 2}\int d^2x\sqrt{-g}\left[{\cal R} {1\over\nabla^2}{\cal R}+
b\,\omega^2\right]\label{uno}\\
&=&{1\over 2}\int d^2x\sqrt{-g}\left[\nabla^\mu\phi\nabla_\mu\phi+
2{\cal R}\phi+b\,\omega^2\right]
\label{local}
\end{eqnarray}
where the  convenient notation has been adopted  
\beq
{\cal R}=\alpha R+\beta \nabla_\mu\omega^\mu\ ,\qquad 
\nabla^2\phi={\cal R}\ ,
\label{due}
\eeq
in order to put action (\ref{uno}) in the compact, and local, form 
(\ref{local}).
$\nabla_\mu$ is the covariant derivative with respect to the Christoffel
symbol, and $\omega_\mu=\epsilon_{ab}e^{b\,\nu}\nabla_\mu e^a{}_\nu$ 
is a Lorentz connection. We adopt a symmetric version
of the induced action because it is easier to work with, but it
can be quickly translated into the asymmetric
form usually appearing in the literature \cite{leut}
\beq
S_{\rm asymm.}=
{1\over 192\pi}\int d^2x\sqrt{-g}\left[R{1\over\nabla^2}\left(\,
{\cal N}R+\Delta {\cal N}\,\nabla_\mu\omega^\mu\right)+
{1\over 2}\,\widehat b\,\omega^2\right]
\label{tre}
\eeq
through the following identification of appropriate parameters
\begin{eqnarray}
\alpha^2+\beta^2&=&{{\cal N}\over 96\pi}\ ,\nonumber\\
\alpha\beta&=&{\Delta {\cal N}\over 192\pi}\ ,\nonumber\\
\widehat b&=&192\pi\left(b-\beta^2\right)\ ,\nonumber\\
{\cal N}&=&n_+ + n_-\ ,\nonumber\\
\Delta {\cal N}&=&n_- - n_+\ ,
\label{quattro}
\end{eqnarray}
where, $n_+\ ,n_-$ are the numbers of left and right chiral fermions
({\it a priori} $n_+\ne n_-$) whose
quantum fluctuations induce the chiral gravitational action. 
The comparison of the actions (\ref{uno}),(\ref{tre}) is possible
because the corresponding non--local parts differ only by a local
$\omega^2$ term, which can be absorbed in a local
counter--term through the redefinition of the free regularization 
parameter $b$. 
The non chiral model corresponds to the choice $ n_+ =n_-$, or
$\beta=0$, and $b=0$. 

The energy--momentum tensor for this model is given by

\begin{eqnarray}
 T_{\mu\nu}&\equiv& 
-{1\over {\rm det}{\bf e}}e_{a\,\mu}{\delta S\over\delta e_a{}^\nu }\nonumber\\
&=&2\alpha\nabla_\mu\nabla_\nu\phi-\nabla_\mu\phi\nabla_\nu\phi
+\beta\,\left(\omega_\mu\nabla_\nu\phi+\omega_\nu\nabla_\mu\phi\right)
\nonumber\\
&&-b\,\omega_\mu\omega_\nu-e^c{}_\mu e_{c\,\nu}
\left(2\alpha\,\nabla^2\phi-{1\over 2}
\nabla^\rho\phi\nabla_\rho\phi+\beta\,\omega^\rho\nabla_\rho\phi-{1\over 2}
b\,\omega^2\right)\nonumber\\
&&-2\beta\left[\epsilon_{\mu\nu}\nabla^2\phi+\epsilon_{\nu\rho}\nabla^\rho
\nabla_\mu\phi\right]+2b \left[\epsilon_{\mu\nu}\nabla\omega+
\epsilon_{\nu\rho}\nabla^\rho\omega_\mu\right]\ .
\label{diciotto}
\end{eqnarray}
It is worth reminding that the \en\ defined with respect to the zweibein
is not a priori symmetric in the indices. This fact will be visible through
the presence  of the Lorentz anomaly \footnote{The non--chiral version
of (\ref{diciotto}) has already been given in \cite{reut}}. 
The conservation laws satisfied by $T_{\mu\nu}$ are 

\begin{eqnarray}
\nabla^\mu T_{\mu\nu}&=&0\ ,\nonumber\\ 
 T&=&2\left(b+\alpha^2\right)R+2\alpha\beta\nabla\omega\ ,\nonumber\\
\epsilon^{\mu\nu}T_{\mu\nu}&=&
-2\alpha\beta R+2\left(b-\beta^2\right)\nabla\omega\ ,
\label{diciannove}
\end{eqnarray}

Equations (\ref{diciannove}) show that we have adopted approach 
 of maintaining maximal possible residual symmetry after integration of 
the matter fields, which produces the anomalous action (\ref{local}). Also,
we have chosen to maintain diffeomorfism invariance at the expenses of the
Lorentz and Weyl symmetry.  Lorentz anomaly is a genuine chiral effect,
and it can be restored only in the non--chiral case
($\beta=0$) by the choice $b=0$. It is worth mentioning that the equations
(\ref{diciannove}) follow directly from equation (\ref{diciotto}) and are,
therefore, {\it off--shell} anomalies.
The essence of the approach described in our paper is to consider induced
anomalous action (2.1) as the ``classical'' action for 2D gravity. 
In other words,
anomalies are crucial to render certain components of the metric (or
zweibein) dynamical, and the anomaly equations {\it are the equations of motion}
for those dynamical components.

Since diffeomorfism invariance is still a
 gauge symmetry of the effective action
(\ref{uno}), we can use this gauge freedom to dispose off some of the 
components of $e^a_\mu$. So, one is faced with the choice of gauge.
We shall study the model in the light--cone gauge, where we can exploit 
the Kac--Moody structure of (chiral) gravity. Moreover, we
shall establish detailed relationships with the parameters
characterizing  the conformal gauge. In order for the reader to be able to 
follow our formulae, we give the decomposition

\beqa
\omega_\mu&=&\partial_\mu L-\epsilon_{\mu\nu}\partial^\nu \varphi
+\bar\omega_\mu\ ,\nonumber\\
\bar\omega_\mu&=&
\epsilon_{ab}\bar e^{b\,\nu}\nabla_\mu \bar e^a{}_\nu\ ,
\nonumber\\
e^{a\,\nu}&=&e^{\varphi/2}\:\left(\begin{array}{cc}\cosh L/2 & -\sinh L/2\\
 -\sinh L/2 & \cosh L/2\end{array}\right)\bar e^{a\,\nu}\ ,
\label{quattrouno}
\eeqa

where, $L$ is the Lorentz degree of freedom, $\varphi$ is the Weyl
degree of freedom, and the rest of the components is 
 contained in barred quantities. So, one can easily switch between 
the two gauges.

\section{Light--cone gauge}
\label{sec:lcone}
\subsection{Residual symmetries of the chiral model}

To introduce the problem, we shall recall some basic results obtained in
the light--cone gauge \cite{smsp}. The metric element is given by
\beq
ds^2=dx^+dx^- +h_{++}dx^+dx^+\ ,
\label{ventidue}
\eeq
and its form is preserved by the following residual symmetry transformation

\begin{eqnarray}
\delta x^+&=&\epsilon^+(x^+)\ ,\nonumber\\
\delta x^-&\equiv&\epsilon^-(x^+,x^-)=
\eta^-(x^+)-x^-\partial_+\epsilon^+(x^+)\ .
\label{duedue}
\eeqa
At this point we would like to stress that diffeomorfisms are
the only remaining symmetry of the induced action due to the presence
of Lorentz and Weyl anomaly (\ref{local}). Therefore, the later two symmetries
are no more an issue, and we shall not be using them.  Diffeomorfisms will be
used to get rid of some of the gauge degrees of freedom leading to the above
choice of the line element. In this case
residual diffeomorfisms reduce to the conformal
transformation in the left--sector, while in the right--sector they represent
a more complex transformation, as written in (\ref{duedue}).\\
The Lorentz degree of freedom $L$, and the metric component $h_{++}$ 
can be defined through the zweibein components given by (\ref{quattrouno}) as 
\beqa
e^L&=&e^{\widehat{+}}{}_+e^-{}_{\widehat{-}}\ ,\nonumber\\
 h_{++}&=&e^{\widehat -}{}_+e^{\widehat+}{}_{+}\ .
\label{duecinque}
\eeqa

while components of the spin connection and of the Christoffel symbols,  are

\begin{eqnarray}
\omega_+&=&\partial_+L+2\partial_-h_{++}\ ,\nonumber\\
\omega_-&=&\partial_-L\ ,\nonumber\\
\Gamma^-{}_{+-}&=&-\Gamma^+{}_{++}=\partial_- h_{++}\ ,\nonumber\\
\Gamma^-{}_{++}&=&\partial_+ h_{++}+2h_{++}\partial_- h_{++}\ .
\label{ventiquattro}
\end{eqnarray}

Lorentz indices  in equation (\ref{duecinque}) and equation (\ref{quattrouno})
(Latin) will be denoted 
by a hat when written in terms of light--cone components to distinguish
them from (~un--hatted~) world (Greek) indices. 
However, hatted indices will not further appear in our formulae.
The dynamical components of the metric and the zweibein vary under the
transformation (\ref{duedue}) as 

\beqa
\delta h_{++}&=&\epsilon^+\partial_+h_{++}+2
h_{++}\partial_+\epsilon^+\nonumber\\
&-&x^-\partial_-h_{++}\partial_+\epsilon^+
-x^-\partial_+^2\epsilon^+ +\eta^-\partial_- h_{++}+\partial_+\eta^-\nonumber\\
\delta L&=& \epsilon^+\partial_+ L-x^-\partial_+\epsilon^+\partial_-L+
\partial_+\epsilon^+ +\eta^-\partial_- L\ .
\label{ventitre}
\end{eqnarray}
We would like to find the generators of the residual symmetries (\ref{duedue})
which should be appropriate combination of the components of \en\ such
that it is  dependent only on the coordinate $x^+$. In order to find such a
combination one can start from the 
symmetry condition (\ref{diciannove}) which in this gauge can be written as
\beq
\partial_-\left(T_{++}-2h_{++}T_{+-}\right)+\partial_+T_{+-}=0 
\eeq
The above equation  has a solution
\beq
T_{++}(x^+,x^-)=\widetilde T_{++}(x^+)+2h_{++}(x^+,x^-)
T_{+-}(x^+)-x^-\partial_+T_{+-}(x^+)
\label{soluz}
\eeq

$T_{++}(x^+,x^-)$ and $T_{+-}(x^+)$ are components of \en\ given 
by (\ref{diciotto}), and $\widetilde T_{++}$ is the generator
of the residual conformal transformation in the left--sector. Its explicit
form will be given later.
We shall also show that the equation of motion
for the induced gravity restrict $T_{+-}$ to be function of $x^+$ alone.  
The components of \en\ coupling to the dynamical degrees of freedom $h_{++}$,
$L$ are
\begin{eqnarray}
 T_{--}(\phi)&=&\left[2(\alpha+\beta)\partial_-^2\phi
-\left(\partial_-\phi\right)^2
+2\beta \omega_-\partial_-\phi-2b\partial_-\omega_-
-b\omega_-^2\right]\ ,\nonumber\\
 \epsilon^{\mu\nu}T_{\mu\nu}&=&-2\alpha\beta R+2\left(b-\beta^2\right)
\nabla\omega\ .
\label{duesei}
\end{eqnarray}

and they lead to the equations of motion given by
   
\begin{eqnarray}
&&\left[\left(\alpha\pm\beta\right)^2
+\left(b-\beta^2\right)\left(1\mp {\alpha\beta\over b-\beta^2}\right)^2\,\right]
\partial_-^3 h_{++}+\alpha\beta\,\partial_-^2 A_+=0\ ,\nonumber\\
&&(b-\beta^2)\,\partial_- A_+=0\ ,\nonumber\\
&&(b-\beta^2)\ne 0\ .
\label{duesette}
\end{eqnarray}
In order to write equations (\ref{duesette}) in the simple looking form we 
have introduced the following redefinitions

\begin{eqnarray}
 A_+&=&D_+L +\left(1-{\alpha\beta\over b-\beta^2}\right)\partial_-h_{++}\ ,
\nonumber\\
D_+ L&=&\partial_+ L-h_{++}\partial_-L\ .
\label{dueotto}
\end{eqnarray}
We would like to mention that the consistency of the equations of motion
(\ref{duesette}), and redefinitions (\ref{dueotto}) require  $b\ne \beta^2$
and, therefore, the previous limit leading to the non--chiral model 
($\beta=0$, $b=0$) cannot be applied anymore. Of course, one could have 
continue to work with the original Lorentz field $L$, but the fields
$h_{++}$, $L$ would not decouple, leading to off--diagonal propagators
and less elegant looking expressions. The non--chiral results
are now simply obtained by dropping all the contribution of the $A_+$ field.
  
The components of the \en\ following from the equation (\ref{diciotto}), and
coupling to the gauge degrees of freedom $h_{--}$, $\varphi$
are 
 
\begin{eqnarray}
 T_{+-}&=&2Q\,\partial_-^2 h_{++}\ ,\nonumber\\
T_{++}-2h_{++}T_{+-}
&=&\left[-\left(\partial_-h_{++}\right)^2+2h_{++}\partial_-^2h_{++}
-2Q\,\partial_-\partial_+h_{++}\right]\nonumber\\
&+&\left[-A_+^2 +2Q_{\rm Lorentz}\,\partial_+A_+\right]
\label{duetrecici}
\eeqa

where, we have rescaled fields as $h_{++}\rightarrow (1/Q_{lc})h_{++}$
\footnote{Since we are considering two different gauge choices, we shall 
denote by a subscript, ``lc''for light--cone,  and ``conf'' for conformal gauge,
all the quantities that a priori
gauge dependent and therefore different. It may, however, turn out 
a posteriori that some of these quantities are actually equal. In that case
we shall drop the appropriate subscript.}, and
$A_+ \rightarrow (1/\sqrt{b-\beta^2})A_+$.
 The constants $Q_{\rm lc}$, $Q_{\rm Lorentz}$ are defined as

\beqa
 Q_{lc}&\equiv& \left[-\left(\alpha\pm\beta\right)^2-
\left(b-\beta^2\right)\left[1\mp {\alpha\beta\over b-\beta^2}\right]^2\,
\right]^{1/2}\ ,\nonumber\\
Q_{\rm Lorentz}&\equiv& \left(b-\beta^2\right)^{1/2}\left(1+{
\alpha\beta\over b-\beta^2}\right)\ ,
\label{dueunoquattro}
\eeqa
and we made use of the identity
\beq
 b\left(1+{\alpha^2\over b-\beta^2}\right)\equiv 
\left(\alpha\pm\beta\right)^2+
\left(b-\beta^2\right)\left[1\mp {\alpha\beta\over b-\beta^2}\right]^2
\label{duqbis}
\eeq
Solutions of the equations of motion (\ref{duesette}) can be written in
terms of four functions $J_i(x^+)$, $\bar J_+(x^+)$
\begin{eqnarray}
 h_{++}(x)&=&J_{++}(x^+)-2x^- J_+(x^+)+\left(x^-\right)^2 J_0(x^+)\ ,
\nonumber\\
A_+(x)&=&\bar J_+(x^+)\ .\nonumber\\
\label{duenove}
\end{eqnarray}
With the help of these solutions one can write the generators of the residual
symmetries as

\begin{eqnarray}
 T_{+-}&=& J_0(x^+)\ ,\nonumber\\
 T_{++}&=&{1\over K}\eta^{ab}J_a J_b
+\partial_+ J_+ +{1\over K_{U(1)}} \bar J_+^2 +\partial_+
\bar J_+\ .
\label{dueunocinque}
\end{eqnarray}
We have further rescaled functions $J_a$ as 
$J_a\rightarrow (1/4Q_{lc})J_a$ and $\bar J_+$ as 
$\bar J_+\rightarrow (1/2Q_{Lorentz})\bar J_+$
 
From the above equation it is visible that the generators are functions
of the coordinate $x^+$ only.
The functions $J_a$, $\bar J_+$ satisfy the Dirac bracket structure 
\beqa
\left\{ J^a(x^+), J^b(x^+)\right\}&=&
-{K\over 2}{\eta^{ab}\over (x^+-y^+)^2}+{f^{ab}{}_c 
J^c(y^+)\over (x^+-y^+)}+ {\rm reg.}   \ ,\nonumber\\
\left\{\bar{J}(x^+),\bar{J}(x^+)\right\}&=&
-{K_{U(1)}\over 2}{1\over (x^+-y^+)^2}+
{\rm reg.}\ ,\nonumber\\
\left\{J^a(x^+),\bar{J}(x^+)\right\}&=&{\rm reg.}\ ,
\label{jj}
\eeqa
and $K$, $K_{U(1)}$ are given by
\beqa
 K&=&- 4Q^2_{lc}\ ,\nonumber\\
 K_{U(1)}&=&-4 Q_{\rm Lorentz}^2\ .
\label{KK}
\eeqa
The above results show that chiral gravity has an underlying 
$SL(2,R)\otimes U(1)$ Kac--Moody structure: 
$f^{ab}{}_c$ and $\eta^{ab}$ being the structure constants and the metric of
the $SL(2,R)$ current algebra\footnote{Our notation can be compared to the
one usually appearing in the literature \cite{Pol3} through the identification
$J_{++}\rightarrow J^{(+)}$, $J_+\rightarrow J^{(0)}$,
$J_0\rightarrow J^{(-)}$.}. 
The explicit form of the $SL(2,R)\otimes U(1)$ generators is
\beqa
l_i^+&=&(x_i^-)^2\partial_{-i}+2\left(\lambda+
{\alpha\beta\over a'-\beta^2}q\right)x_{-i}\ ,\nonumber\\
l_i^0&=&x_i^-\partial_{-i}+\left(\lambda+
{\alpha\beta\over a'-\beta^2}q\right)\ ,\nonumber\\
l_i^-&=&\partial_{-i}\ ,\nonumber\\
{\bar l}_i^0&=&q\ .
\label{elles}
\eeqa
where, $\lambda$ is a scale weight in the right sector of a primary field
$\Phi$, and $q$ is its Lorentz weight. 
So far, we have described the ``classical'' symmetries of the model. 
Quantizing the model means calculating gravitational contribution to 
the ``classical'' action (\ref{local}). This can be done in two ways: 
i) by direct Feynman
graph calculation \cite{oz2}, which is interesting but tedious due to the 
non--local form of the action, or ii) requiring
the persistence of the classical symmetries (\ref{duedue}) at the quantum 
level \cite{Pol6}. This amounts to the 
requirement of the weakly vanishing of the 
appropriate generators $\widetilde T_{++}$ and  $T_{+-}$. We shall follow the
second approach. The quantum analogue of the classical Sugawara form 
(\ref{dueunocinque}) is well known \cite{fre} and is given by  

\begin{eqnarray}
 T_{+-}&=& J_0(x^+)\ ,\nonumber\\
\widetilde T_{++}&=&{1\over K+2}\eta^{ab}:J_a J_b:
+\partial_+ J_+ +{1\over K_{U(1)}}: \bar J_+^2: +\partial_+
\bar J_+\ .
\label{tii}
\end{eqnarray}
where, ``~$+2$~'' is the Casimir of the $SL(2,R)$ and ``~dots~'' mean 
appropriate normal ordering of the currents.

It is further possible to show that the quantum generators of the residual 
symmetry transformations satisfy the following OPE's
\beqa
\widetilde T_{++}(x) \widetilde T_{++}(y)
&=&{3K/(K+2)-6K +1-6K_{U(1)}\over 2(x^+-y^+)^4}+\nonumber\\
&&{2 \widetilde T_{++}(y)\over (x^+-y^+)^2}+{\partial_+
\widetilde T_{++}(y)\over x^+-y^+}+{\rm reg.}\ ,\nonumber\\
&{}&\nonumber\\
\widetilde T_{++}(x)T_{+-}(y)&=&{2T_{+-}(y)\over (x^+-y^+)^2}+
{\partial_+T_{+-}(y)\over x^+-y^+}+{\rm reg.}\ ,\nonumber\\
&{}&\nonumber\\
T_{+-}(x)T_{+-}(y)&=&{\rm reg.}
\ .\label{dueunootto}
\eeqa
It can be seen from (\ref{dueunootto}) that the weakly vanishing of 
$\widetilde T_{++}$ and $T_{+-}$ is possible only if the total 
central charge, including
matter and ghost contributions, is vanishing. This condition leads to the 
renormalized central charge through the equation 
\beqa
&&n_+ -28 +c_{\rm grav.}=0\ ,\nonumber\\
&&c_{\rm grav.}={3K\over K+2}-6K +1+ 24 Q_{Lorentz}^2\ .
\label{dueunonove}
\eeqa
where, $n_+$  and $-28$ are respectively the chiral matter
and the ghost contribution resulting from gauge fixing (\ref{ventidue}).
Equations (\ref{dueunonove}) give the renormalized Kac-Moody $SL(2,R)$
central charge
\beq
K+2={(n_+ -12+24 Q_{Lorentz}^2)\pm 
\sqrt{(n_+ +24 Q_{Lorentz}^2)(n_+ -24+ 24 Q_{Lorentz}^2)}
\over 12}
\label{dueduezero}
\eeq
From the equation (\ref{dueduezero}) follows
that the renormalized  Kac-Moody $SL(2,R)$ central charge is dependent on a
free parameter, through the $Q_{Lorentz}$,
 that can be adjusted at will to make a square root in
(\ref{dueduezero}) real for any value of $n_+$ (see discussion
in Section(\ref{restrict})). Therefore, {\it contrary to the
non--chiral gravity, in chiral models
 there are no forbidden regions corresponding to complex values of
the central charge} since it can always be made real. 
This is due to the presence of the Lorentz degree of
freedom which acts as an additional ``~matter~'' contribution introducing, 
at the same time, a new dependence on a free parameter. Such a parameter has
no fixed value in chiral models since there is no value that can remove 
the Lorentz anomaly. 

\subsection{Generalization and extension to supersymmetric models} 

In the previous subsection we have explained in detail the approach
to the induced  chiral gravity models. Prescriptions to
obtain non--chiral results were given as well. 
Supersymmetric versions of the same models could be studied along the same
lines but one has to have a detailed knowledge of the appropriate
superspace structure model by model. Instead, we would like to
present a simple, generalized, approach reproducing all the known results
with little effort. 
It is based on the following observation. In order to maintain
residual symmetries at the quantum level, in any model, one has to require
the weakly vanishing of the residual symmetry generators
leading to the vanishing of the total Virasoro central charge. 
 
\beq
c_{\rm matt.} +c_{\rm grav.}^{\rm lc}  +c_{\rm ghosts}^{\rm lc}=0\ ,
\label{cuno}
\eeq
This is the formula which is subject to a simple generalization.
$c_{\rm grav.}$ can be obtained from the equation (\ref{tii})
for an arbitrary group as
\beq
c_{\rm grav.}^{\rm lc}={K{\rm dim}(G)\over K+C_v}-6K \ .
\label{219}
\eeq
where,
${\rm dim}(G)$ is the dimension of the appropriate $SL(2,R)$ group (or its 
supersymmetric version),  $C_v$ is the corresponding Casimir,
and $K$ is the renormalized Kac--Moody central charge of the same group.
The generality of this expression stems from the fact that differences
among various models are all contained within $c_{\rm matt.}$, 
$c_{\rm ghosts}^{\rm lc}$, while (\ref{219}) 
takes into account the structure of the light--cone gauge
group of residual symmetry. It works both for non chiral and chiral models
since in the latter case the Lorentz degree of freedom acts as additional
``matter'' field, as far as $SL(2,R)$ group is concerned,
 and it will be contained in $c_{\rm matt.}$. It will be 
explicitly shown later on. \\
Solving the algebraic quadratic equation
(\ref{219}) with respect to $K$ 
gives the gravitationally renormalized  Kac--Moody central charge. 
  If we denote the two roots $\kappa_\pm= K_\pm +C_v$ they will satisfy 
	the following relation
	\beqa
        \kappa_+\kappa_-&=&a^2\label{102}\\
        \kappa_+ +\kappa_- &=& 2a-m Q^{2}_{lc}/2 \label{12}
	\eeqa
	We have introduced a parameter $m$ to explicitly stress 
	the difference among various supersymmetric models: $m=1$
	for $N=0,1$, and $m=2$ for $N=2$. 
	From equations (\ref{cuno}), (\ref{219})
  	the parameters $a$ and $Q_{lc}$ are found to be
	
	\beqa
	a^2&=&C_v{\rm dim}(G)/6\label{4.5}\\
        Q_{lc}^2&=&-{1\over 3m}\left[c_{\rm matt.}+c_{\rm ghosts}^{\rm lc}+
	\left(\sqrt{6C_v}-
	\sqrt{{\rm dim}(G)}\,\right)^2\,\right]\nonumber\\ 
	\label{4.6}
	\eeqa

	In this way  we
	get a simple evaluation of these parameters  {\it in terms}
	of the characteristics of the residual symmetry group
	in the light--cone gauge. At this point it may be 
	instructive to calculate these parameters for various models.
	First of all, let us calculate Casimir operators for various
	supersymmetric versions of the $SL(2,R)$ group. Its definition is
	\beq
	C_v\delta_a{}^b=f_{acd}f^{bcd}
	\eeq
	where, $f_{abc}$ are anti--symmetric structure constants of the 
	appropriate  $SL(2,R)$ group. Using the values of these constants
	for $N=0,1,2 $   one can
	find the general formula for the Casimir $C_v=2-N/2$. For  
	$dim(G)=\# \hbox{(bosonic generators)}-
	\# \hbox{(fermionic generators)}$ we have values $3,1,0$ respectively.\\
	In order to calculate ghost contributions, let us look at the 
	constraints of the residual symmetry generators. For $N=0$
	they are $\widetilde T_{++}$ and $T_{+-}$ leading to the
	ghosts $(b_{++}, c_-)$ and $(b, c_+)$, with contributions $-26$, 
	$-2$ \cite{Pol4}.
	
	In the case $N=1$ the constraints are  
	$(\widetilde T_{++}, G_{1/2\, +})$, $(T_{+-},J_{1/2})$. The
	appropriate ghosts are $(b_{++}, c_-)$, $(b_{1/2\, +}, c_{1/2})$,
	 $(b, c_+)$, and the single
	ghost $b_{1/2}$ with contributions $-26$, $11$, $-2$, $-1/2$
	\cite{Pol5}.\\
	Subscript $1/2$ refers to the fermionic partner of the bosonic
	coordinate $+$.\\
	In the case $N=2$ the only difference with respect to the $N=1$ case
	is the doubling of the ghost and matter contribution
	 due to additional supersymmetry \cite{Pol5}.
	 Therefore, the counting of ghost contributions
	is $-26+2\times 11-2\times 2-2\times (1/2)$. 
		
	\medskip
	\begin{table}[hbt]
	\smallskip
	\begin{tabular}{|p{3cm}|p{3cm}|p{3cm}|p{3cm}|}
	\hline\hline
	\phantom{$N=0$} &\smallskip\hfill $ N=0$\hfill\smallskip &
	\smallskip\hfill  $ N=1$ \hfill\smallskip& \smallskip\hfill $ N=2 $
	\hfill\smallskip\\
	\hline
	 \smallskip\hfill $ dim(G)$ \hfill\smallskip &
	\smallskip\hfill $3$ \hfill\smallskip &
	\smallskip\hfill $ 1$ \hfill\smallskip &
	\smallskip\hfill $0$ \hfill\smallskip\\
	\hline
	\smallskip\hfill $C_v$ \hfill\smallskip &
	\smallskip\hfill $2$ \hfill\smallskip &
	\smallskip\hfill $ 3/2$ \hfill\smallskip & \smallskip\hfill $1$ 
	\hfill\smallskip\\
	\hline
	\smallskip\hfill    $a$  \hfill\smallskip &
	\smallskip\hfill  $1$ \hfill\smallskip &
	\smallskip\hfill  $1/2$ \hfill\smallskip &\smallskip\hfill  $0$
	\hfill\smallskip\\
	\hline
	\smallskip\hfill $c_{matt}$ \hfill\smallskip &
	 \smallskip\hfill $d$ \hfill\smallskip &
	\smallskip\hfill  $3d/2$ \hfill\smallskip &
	\smallskip\hfill $3d$ \hfill\smallskip\\
	\hline
	 \smallskip\hfill $c_{ghost}^{lc}$ \hfill\smallskip &
	\smallskip\hfill  $-28$ \hfill\smallskip &
	\smallskip\hfill  $-35/2$ \hfill\smallskip &
	\smallskip\hfill  $-9$ \hfill\smallskip\\
	\hline
	\smallskip\hfill $Q^2_{lc}$ \hfill\smallskip &
	\smallskip\hfill $ (25-d)/3$ \hfill\smallskip &
	\smallskip\hfill  $ (9-d)/2$ \hfill\smallskip &
	\smallskip\hfill $(1-d)/2$ \hfill\smallskip\\
	\hline\hline
	\end{tabular}
	\caption{
	We display values of various parameters entering formulae
	(\protect\ref{4.6}) in the case of non--chiral models. 
	Parameter $d$ refers to the number of scalar components,
	while left/right fermionic components are described by $n_{+/-}$.
	Supersymmetry imposes $n_+=d$ 
	in the count of matter contribution,
	e.g.  $d+n_+/2=3d/2$. }
	\label{table1}
	\end{table}
	\medskip
	
	\begin{table}[htb]
	\smallskip
	\begin{center}
        \begin{tabular}{|p{3cm}|p{7cm}|}
        \hline\hline
        \phantom{$N=0$} &\smallskip \hfill  $\kappa_\mp$\hfill\smallskip \\
        \hline
        \smallskip \hfill  $N=0$  \hfill\smallskip
        &\smallskip \hfill $\left(\, d-13\pm\sqrt{(d-1)(d-25)}\,\right)/12$
        \hfill\smallskip   \\
        \hline
	\smallskip \hfill $N=1$  \hfill\smallskip
        & \smallskip\hfill $\left(\, d-5\pm\sqrt{(d-1)(d-9)}\,\right)/8$
        \hfill\smallskip    \\
        \hline
        \smallskip \hfill $N=2$ \hfill\smallskip
        &\smallskip \hfill $\left[\left(d-1\right)\pm \left(1-d\right)\right]
	/4$ 
	\hfill\smallskip   \\
	\hline\hline
         \end{tabular}
	\caption{ We display the values of the renormalized Kac--Moody central
	charge as defined by (\protect\ref{12}) with the help of the result
	in the previous Table. Notice that in the $N=2$ case the quadratic 
	equation
	reduces to the linear one, and the absence of the square root means
	the absence of the complex values of the renormalized Kac--Moody 
	central charge. This is due to the property $ dim(G)=0$ in this case.}
	\end{center}
	\label{table2}
	\end{table}
	\medskip
	
	The results for various models are presented in 
	Table 1,
	while the values of the renormalized Kac--Moody central charge are
	given in Table 2

	 One of the nice features of the light--cone gauge is that
	one  can also perform genuine perturbative
	calculations, due to the presence of a well defined regulator. 
	This calculation serve as a check of the closed formulae displayed
	in the Table 2, as well as enable access to the 
	``large $d$'' region, where stringy character of the induced gravity
	model is relevant. Perturbative calculations \cite{oz2} are, however, 
	complicated due to the non--local form of the effective action.
	Certain shortcuts can be exploited at the one--loop level, but
	the complication remains to higher level. 

	We would like to give a
	simple determination of the perturbative results (up to one--loop)
	based on our general equations (\ref{12}). Classical Kac--Moody
	central charge $K_{\rm cl.}$ is a coupling constant in the classical
	action (\ref{tre}) as can be seen from (\ref{KK}). From (\ref{12})
	one can see that only $\kappa_+$ has a proper classical limit 
	$c_{matt}>>1$, which is $\kappa_+ \equiv K_{cl.}=c_{matt}/6$,
	while $\kappa_- \sim O(1/c_{matt})$. 
	One can perform perturbative calculations
	in $1/K_{\rm cl.}$, leading to the renormalized Kac--Moody  central 
	charge $K=Z_K K_{\rm cl.}$. Equations (\ref{12}),
	 (\ref{4.6}) allow a simple determination of the counterterm
	$Z_K $, which is found to be \cite{oz2} 
	\beq
	Z_K=1+{1\over c_{\rm matt.}}\left(c_{\rm ghost}^{\rm lc}+{\rm dim}(G)
	\right)-{6\kappa_-\over c_{\rm matt.}}
	\label{zren}
	\eeq
	From (\ref{zren}) one can see that all the higher loops contributions
	are hidden within $\kappa_-$. 
	\medskip
	\begin{table}[hbt]
	\begin{center}
        \begin{tabular}{|p{3cm}|p{3cm}|}
        \hline\hline
        \phantom{$N=0$} &\smallskip \hfill  $Z_K$\hfill\smallskip \\
        \hline
        \smallskip \hfill $ N=0$  \hfill\smallskip
        &\smallskip \hfill $1-(25/d)$
        \hfill\smallskip   \\
        \hline
        \smallskip \hfill $ N=1$  \hfill\smallskip
        & \smallskip\hfill $1-(11/d)$
        \hfill\smallskip    \\
        \hline
        \smallskip \hfill $N=2$ \hfill\smallskip
        &\smallskip \hfill $1-(3/d)$ \hfill\smallskip   \\
        \hline\hline
 	\end{tabular}
	\caption{We display the one loop results for the Kac--Moody central
	charge counterterm following from (\protect\ref{zren}). 
	This results can be
	also obtained expanding results in Table 2 
	in powers of $1/d$.
	Notice that $N=2$ case has only one--loop contributions. This is
	due to the fact that $\kappa_- =0$ in this case. }
	\end{center}
	\label{table3}
	\end{table}	
	\medskip

Another formula, that can be easily generalized and leads to
a common treatment of various induced (super) gravity models, is the one
following from the imposition of the weakly vanishing condition on the physical
states within the BRST quantization procedure in the light--cone gauge.
$\widetilde T_{++}$ is the generator of the conformal 
transformations in the left moving sector, and is expressed through the 
Sugawara construction in terms of the $SL(2,R)$ currents (or their
supersymmetrized version) as given by (\ref{dueunocinque}). 
One of these currents, i.e. $J^0$, describes, on the other hand, the scale 
transformation with weight $\lambda$ in the right moving sector as visible from
(\ref{elles}). On general grounds \cite{conf}, scale weight of the primary 
field, with respect to the scale transformations in the left sector, is
$\Delta^{\rm left}$, while with respect to the right sector it is 
$\Delta^{\rm right}$.
Therefore, in this particular case $\lambda\sim \Delta^{\rm right}$. 
	 Applying the generator 
	of the conformal transformations in the left moving
	sector  on a primary field, and exploiting (\ref{dueunocinque})
	 one finds its conformal weight\cite{Pol6} to be
	\beq
	\Delta^{\rm left}=
	{\lambda(\lambda+1-N/2)\over K+C_v}-\lambda
	\label{4.2bis}\ ,
	\eeq
	BRST analysis of the constraints of the residual symmetry has been
	performed in the light--cone gauge, and appropriate definitions of the 
	physical states has been given in \cite{ung}, \cite{11}. Weakly 
	vanishing of the generators of residual symmetry lead to the equation

	\beq
        \bar a-\Delta_0
	={\lambda_\beta(\lambda_\beta+ a)\over K+C_v}-\lambda_\beta
        \ ,\label{dueunodieci}
        \eeq
	where, $\bar a$ is the normal ordering ambiguity parameter
        of the BRST formalism and $\Delta_0$ is the scaling weight 
	of the matter field. 
	Equation (\ref{4.2bis}) give the value of the parameter $a$
	for various supersymmetric models as $ a=1-N/2$. Comparison to the
	Table 1 shows that the explicit
	values of $a$, as defined above, are equal to the ones defined by
	(\ref{4.5}). Furthermore, we shall show later on that
	 $ a=\bar a$. Accordingly, from now on we shall be using only one 
	symbol ``$a$''.\\ 
	At this point we would like to mention that we are still
	working in full generality, in the sense that the results will
	apply both to chiral and non chiral (supersymmetric) models, since
	the presence of the (super)Lorentz field will be contained within
	$\Delta_0$ being just another ``matter'' field. An explicit example 
	will be given later.\\ 
	The  subscript $\beta$ in $\lambda$
	refers to the solution of the above equation
	in the presence of matter $\Delta_0\ne 0$. When $\Delta_0=0$ then 
	$\beta\rightarrow \alpha$. The subscripts $\alpha$,$\beta$ have been 
	introduced in order to make comparison to the conformal gauge results
	(which will follow) more transparent.\\
	The roots of the algebraic quadratic equation 
	(\ref{dueunodieci}) satisfy 
	\beqa
        &&\lambda_{\beta_+} \lambda_{\beta_-}=(K+C_v)(\Delta_0-a)\nonumber\\
        &&\lambda_{\beta_+}+ \lambda_{\beta_-}=K+C_v-a
        \label{1.2}
        \eeqa
        The vacua will correspond to $\Delta_0=0$, or 
	$\beta\rightarrow\alpha$, and are characterized by the scaling weights
        \beqa
        &&\lambda_{\alpha_+}=-a\nonumber\\
        &&\lambda_{\alpha_-}=K+C_v
        \label{vuoti}
	\eeqa
	Equations (\ref{vuoti}) show that there are always two distinct
         vacua described by the above $SL(2,R)$ weights, which are not,
        in general, projectively invariant. \\
	At this point we can establish relations between the renormalized
	Kac--Moody central charge and the above vacua as
	\beq
        \kappa_\pm=  {\lambda_{\alpha_\mp}^2\over \lambda_{\alpha_-}} \ .
        \label{kappaeq}
	\eeq

	\section{Connection between the light--cone and conformal gauge}

	In the conformal gauge the metric element takes on the form
	\beq
	ds^2=e^\varphi dx^+ dx^-
	\eeq
	Once this coordinate mesh is set over spacetime, 
	the dynamics of the $\varphi$ is dictated by the Liouville 
	action,  and a residual invariance is present as a
	 conformal symmetry in both left and right moving sectors.
	The gravitational interaction is represented by the Liouville potential
	with a coupling constant given by the non zero cosmological constant.
	Perturbative calculations are ambiguous
	in this gauge due to the lack of an appropriate regulator, as well
	as of a ``small'' parameter justifying a perturbative expansion
	\cite{dock}. 
	As an alternative to perturbative calculations,
	the DDK approach \cite{distl}, as well as
	the cocycle quantization \cite{li}, has been recently offered. 
	The basic underlying idea in this approach is to replace the
	original, intractable, functional measure depending from
	the Weyl degree of freedom, with an invariant measure
	with respect to a background metric $\widehat g_{ab}$. 
	The original metric is split as 
	$\displaystyle{g_{ab}=\widehat g_{ab} e^{\alpha\varphi}}$.
	The Jacobian of the transformation is {\it assumed} to be local and
	of the same form as the original Liouville action, but with two
	unknown coefficients. One of them is the background charge $Q_{conf}$
	and the other is the gravitational renormalization parameter $\alpha$
	of the Liouville potential $\displaystyle{e^{\alpha\varphi}}$.
	The two free parameters are then fixed by the requirement of scale
	invariance with respect to the {\it background metric} 
	$\widehat g_{ab}$. The above mentioned invariance leads to the 
	conformal gauge equivalent of the light--cone equation (\ref{cuno}), 
	and is given by
	\beq
	c_{\rm matt.}+c_{\rm ghost}^{\rm conf.}+c_{\rm grav.}^{\rm conf.}=0
	\label{ccq}
	\eeq
	where $c_{\rm grav.}^{\rm conf.}$ is given by 	 
	\beq
	c_{\rm grav.}^{\rm conf.}=m\left(1+3Q^2_{conf}\right)+{N\over 2}
	\label{cm}
	\eeq	  
	Parameter $m$ enters the equation (\ref{cm}) due to the additive
	character of the central charge, and the fact that $N=2$ supersymmetry
	corresponds to doubling of the number of Weyl degrees of freedom.
	$N/2$ is the contribution of the corresponding superpartners.  
	The parameter $Q_{conf}$ obtained from the
	equation (\ref{ccq}) is given by 
	\beq
	Q^2_{conf}=-{1\over 3m}\left[c_{matt}+c_{\rm ghost}^{\rm conf.}+
	m+{N\over 2}\right]
	\label{qconf}
	\eeq
	One could prove by explicit calculations that the $Q_{conf}$ given by
	(\ref{qconf}) and $Q_{lc}$ given by (\ref{4.6}) are identical.
	We shall give proof of this statement on general grounds later on.
	
	The effect of Weyl dynamics in the conformal gauge is that the 
	various classical operators 
	will receive gravitational dressing described by
	the parameter $\beta$ in the  Liouville potential 
	$\displaystyle{\Phi e^{\beta\varphi}}$, where $\Phi$ represents
	conformal matter field. 
	As we have mentioned earlier, the essence of the DDK approach consists
	of requiring the invariance under scale transformation with respect
	to the background metric $\hat g_{\mu\nu}$. Applying this condition to
	the Liouville potential in the presence of matter puts constraints 
	on the total scale weights of such potential. These constraints can be
	determined in the general way by considering the scale weight of the
	super integration measure $d^2x\, d\theta^+d\theta^-$. 
	Keeping in mind that 
	the scale weight of the superspace parameters is $1/2$, one obtains
	the scale weight of the integration measure as $(-1+N/2, -1+N/2)$,
	which then gives the scale weight of the Liouville potential
	$(1-N/2, 1-N/2)\equiv (\bar a,\bar a)$. As stated earlier, this
	shows the equality $\bar a =a$, where $a$ is defined in 
	(\ref{dueunodieci}). 
	If we denote $\Delta_0$ the scale weight of the spinless matter field
	$\Phi$, then the scale weight of the Liouville potential is
	
	\beq
	a -\Delta_0=-{m\over 2}\beta(\beta +Q_{conf})
	\label{5.2}
	\eeq
	Equivalent expression in the light--cone gauge is
	\beq
         a-\Delta_0
	={\lambda_\beta(\lambda_\beta+ a)\over K+C_v}-\lambda_\beta
        \label{2110}
        \eeq
 
	Confronting the above equation in case $\beta=\alpha$, 
	e.g. $\Delta_0=0$, we obtain
\beq
-{m\over 2}\alpha^2\left(1+ {Q_{conf}\over \alpha}\right)=
{\lambda^2_\alpha\over
K+C_v}\left(1-{K+C_v-a\over\lambda_\alpha }\right)
\eeq
from which we deduce the relation among various parameters in the two gauges
\beq
-{m\over 2}\alpha^2= {\lambda^2_\alpha\over K+C_v}
\label{alfa}
\eeq
and
\beq
{Q_{conf}\over\alpha}=-{1\over \lambda_\alpha} (K+C_v-a)
\label{qconf2}
\eeq
From equations (\ref{alfa}), (\ref{qconf2}) 
one can prove the following relations
\beq
Q^2_{conf}=-{2\over m}{\left(K+C_v-a\right)^2\over K+C_v}
\label{qqconf}
\eeq
Further, exploiting relation (\ref{1.2}) and the definitions of $\kappa_\pm$
one obtains the following relation
\beq
 \kappa_+ +\kappa_- = 2a-m Q^{2}_{conf}/2
\eeq
which proves the identity between $Q_{conf}$ and $Q_{lc}$ on general grounds.
Comparing the explicit expression (\ref{qconf2}) and (\ref{4.6})
	we get a particularly simple relation between ghost contributions 
	in the two gauges
	\beq
	c_{\rm ghost}^{\rm conf.}=c_{\rm ghost}^{\rm lc}+\left(\sqrt{6C_v}
	-\sqrt{dim(G)}\,\right)^2-m-{N\over 2}
	\label{cvari}
	\eeq

 Further identification of parameters in the two gauges can be obtained
 considering the equation (\ref{2110}) in the presence of matter 
($\Delta_0\ne 0$). In this case, one  can rewrite equation (\ref{2110}),
with the help of (\ref{alfa}), (\ref{qconf2}),  as
\beq
-{\beta\over\alpha}\lambda_\alpha + {1\over K+C_v}{\beta\over\alpha}
\lambda_\alpha\left(a+{\beta\over\alpha}\lambda_\alpha\right)=
{\lambda_\beta(\lambda_\beta +a)\over K+C_v}-\lambda_\beta
\eeq
which gives the following relations among various parameters
	\beq
	\lambda_\beta={\beta\over\alpha}\,\lambda_\alpha\ ,\label{5.5}
	\eeq
	
	\beq
	{Q\over\alpha_\pm}=-1-{\lambda_{\alpha_{\mp}}\over 
	\lambda_{\alpha_{\pm}}}\ ,\label{5.7}
	\eeq

	\beq
        \alpha_\pm^2=-{2\lambda_{\alpha_\pm}^2\over
        m\lambda_{\alpha_-}}\ , \label{5.8}
        \eeq
	Equations (\ref{5.7}), (\ref{5.8}) are the same as (\ref{alfa})
	and (\ref{qconf2}), which serves as a consistency check. 
	We have dropped the subscripts $_{conf}$,
	$_{lc}$, on the parameter $Q$ in view of the identity $Q_{conf}\equiv
	Q_{lc}$, as proven above. 

	\subsection{Particularity of the chiral models}

	Chiral models are characterized by the presence of the dynamical
	Lorentz degree of freedom, which enters through the
	zweibein as the fundamental ingredient of the theory. The conformal 
	gauge expression is 
	\beq
	e_\mu{}^{\widehat\pm}=\exp\left({\varphi\pm L\over 2}\right)
	\,\delta_\mu{}^{\widehat\pm}
	\eeq
	However, as far as DDK approach is concerned, the functional measure
	of the path integral for the dynamical fields is only dependent on the
	determinant of the zweibein and, therefore, independent of the Lorentz
	degree of freedom. So, prescription of defining a modified (proper) 
	measure, will go unchanged in the chiral case due to the fact that the
	line element is Lorentz independent \cite{mp}. 
	In this respect, Lorentz field
	appears to be just additional matter field in the theory.
	 Its scale weight
	can be obtained considering its OPE with the generator of the
	conformal transformation as given by the component $ T_{\pm\pm} $
	of the \en\ following from (\ref{diciotto}), which are

	\beqa
	T_{++}&=&(\partial_+\varphi)^2 - 2Q\,\partial_+^2\varphi+
	(\partial_+\chi)^2 - 2Q_{\rm Lorentz}^{\rm left}\,
	\partial_+^2\chi\\
		T_{--}&=&(\partial_-\varphi)^2 - 2Q\,\partial_-^2\varphi+
	(\partial_-\chi)^2 - 2Q_{\rm Lorentz}^{\rm right}\,
	\partial_-^2\chi
	\eeqa

	where, we have used a redefinition of the Lorentz field as
	$\chi=L+\alpha\beta\varphi/(b-\beta^2)$. 
	The presence of $Q_{\rm Lorentz}^{\rm left/right}$ 
	indicates that there is
	a background Lorentz charge in the theory, and it is different in the
	left/right sector as can be seen from the expressions  
	\beqa
	Q_{\rm Lorentz}^{\rm left}&\equiv&
	\left(b-\beta^2\right)^{1/2}\left({
	\alpha\beta\over b-\beta^2}+1\right)\\
	 Q_{\rm Lorentz}^{\rm right}&\equiv&
        \left(b-\beta^2\right)^{1/2}\left({
        \alpha\beta\over b-\beta^2}-1\right)
	\eeqa

	Lorentz field will enter the Liouville potential in the exponential 
	form as $e^{\kappa\chi}$, $e^{\tilde\kappa\chi}$ where $\kappa$, 
	$\tilde\kappa$ are holomorphic (left), and anti--holomorphic (right)
	momenta of the Lorentz field. They are introduced due to the fact
	the Lorentz field carries spin, i.e. couples to the spin connection
	and therefore one has to distinguish between left and right components.
	Its scale dimension, in the left sector, is given by 
		
	\beq
	\Delta_{\rm Lorentz}^{\rm left}=
	-{m\over 2}\kappa(\kappa+Q_{\rm Lorentz}^{\rm left})
	\label{deltalor}
	\eeq
	and an analogous equation is valid in the right sector 
	substituting $\kappa\rightarrow \tilde\kappa$ and 
	$Q_{\rm Lorentz}^{\rm left}\rightarrow Q_{\rm Lorentz}^{\rm right}$.
	The presence of spin is also visible through the fact 
	$Q_{\rm Lorentz}^{\rm left}\ne Q_{\rm Lorentz}^{\rm right}$.
	In fact,  from the definition of 
	spin $s= \Delta^{\rm right}-\Delta^{\rm left}$, 
	and from the above equations one finds
  $s=(\kappa m/2)(Q_{\rm Lorentz}^{\rm left}-Q_{\rm Lorentz}^{\rm right})\ne 0$
	We have assumed simplified case of $\tilde\kappa=\kappa$ which gives
	a simple looking formula. It can be shown, in general, that the 
	dressing operators with spin are necessary in chiral theories in
	order to satisfy momentum conservation rules in the presence
	of the background charges and obtain non--trivial correlation
	functions. Such operators are discussed in \cite{mp}
	and we refer reader for details to it. We repeat just notions
	necessary for our purpose. Given a matter operator of conformal
	weight $(\Delta^{\rm left}$,  $\Delta^{\rm right})$, it will acquire
	gravitational dressing of the form $e^{\alpha\varphi} e^{\kappa\chi}$
	in the left sector (analogous situation in the right sector).
	We shall assume that the Weyl field $\varphi$ carries no spin as in
	the non chiral case in order to be able to reproduce a local operator
	analog of the cosmological constant. In order to make a dressed
	 operator of $(a,a)$ scale weight, it has to satisfy the analog  	
	of the equation (\ref{5.2}) in the chiral case 
		\beq
	a -\Delta_0^{\rm left}-\Delta_{\rm Lorentz}^{\rm left}=
	-{m\over 2}\beta(\beta +Q_{conf})
	\label{5.2bis}
	\eeq
	Analogous extension of the light--cone equation (\ref{2110}) is 
	\beq
         a-\Delta_0
	={\lambda_\beta(\lambda_\beta+ a)\over K+C_v}-\lambda_\beta
	+{q^2_\beta\over K_{U(1)}}-q_\beta
        \label{21100}
        \eeq
	Comparing these two equations in the same way  followed in the 
	non--chiral case
	we obtain the relations (\ref{5.5}), (\ref{5.7}), (\ref{5.8}),
	and, in addition, relations
	between the appropriate parameters characterizing the Lorentz field
	in the two gauges 
	\beq
	{Q_{\rm Lorentz}\over k_\alpha}=-{K_{U(1)}\over q_\alpha}\ ,
	\eeq	
	\beqa
	&&q_\beta={k_\beta\over k_\alpha}\, q_\alpha\\
	&&k_\beta^2=-{2q_\beta^2\over m K_{U(1)}}
	\label{qek}
	\eeqa
	where, $k_\beta$ is the gravitational dressing parameter of the
	Lorentz field in presence of matter. In principle, one can consider
	quantum operators with scale weight $(a^{\rm left}, a^{\rm right})$
	 with $a^{\rm left}\ne a^{\rm right}$
	which are quantum analogs of classical operators involving zweibein
	and its gravitational covariant derivatives. Therefore, in chiral
	theories the choice of dressing operators is richer than in 
	non--chiral case. However, these are necessarily non--local operators. 
	In order to reproduce a local operator (cosmological
	constant) as in non--chiral case absence of matter requires $\Delta_0=0$
	and $\Delta_{\rm Lorentz}=0$. This gives values $\kappa_\alpha=0$, and  
	$\kappa_\alpha=-Q_{\rm Lorentz}$ in the conformal gauge, or 
	$q_\alpha=0$ and $q_\alpha=K_{U(1)}$. Therefore, we see that in the
	chiral gravity vacuum states are both scale and Lorentz non--invariant.
	 
	\subsection{Derivation of KPZ equation}

	The main effect of gravitational dynamics is to renormalize
	flat space quantities. Among them a particular interest is in
	calculating gravitational effects on the naive
        flat space scaling dimensions of various fields in the theory 
	\cite{Pol6}. One approach is to notice that the gravitational
	effects will violate conformal block decomposition of the
	multipoint functions. Exploiting differential equations that
	these functions satisfy in conformal field theory, one can obtain
	the above mentioned renormalized values of the scaling dimensions.
	This works fine as long as one works with projectively invariant
	vacua. Unfortunately, models described in this paper have 
	non--invariant vacua and only hand--waving arguments \cite{ung}
	lead to the determination of the scaling dimensions. 
	On the other hand, in the DDK 
	approach a variation of the area, i.e. cosmological constant,
	with respect to the background scale transformations enable one
	to determine a renormalized scaling dimension in terms of parameters
	$\alpha$, $\beta$ \cite{ddk2}. At this point we would like to
	mention that the DDK derivation of the renormalized scaling dimension
	would go unchanged in the presence of the Lorentz field. This is
	due to the fact that the background scale transformation affects
	{\it only Weyl degree of freedom}, and leaves Lorentz field 
	unchanged. That is way it is considered just another ``matter'' field.

	Therefore, starting from the definition of the gravitational scaling
	weight, as obtained in the conformal gauge, 
	
	\beq
        \Delta =a\left(1-{\beta\over\alpha}\right)
        \label{1.3}
        \eeq
	and exploiting relation (\ref{5.5}) one can write the above
	weight in the light--cone gauge as
	
	\beq
        \Delta =a\left(1-{\lambda_\beta\over\lambda_\alpha}\right)
        \label{1.30}
	\eeq
	Equation (\ref{1.30}) indicate that one is actually referring to the
	renormalized scaling weight in the right sector 
	$\Delta=\Delta^{\rm right}$ since this one is related to the scale
	 weight $\lambda$. Such a specification was not important 
	in the non--chiral case since matter fields are spinless. As we have 
	shown in the chiral case, we have to require
	$\Delta^{\rm left}\ne \Delta^{\rm right}$.
	 Inserting (\ref{1.30}) into (\ref{dueunodieci}) leads to the
	KPZ equation for the non--chiral models 
	
	\beq
        \Delta -\Delta_0=\kappa
        {\Delta\over a}\left({\Delta\over a} -1\right)\label{1.4}
        \eeq

	In the case of chiral models, the above equation becomes

	\beq
        \Delta^{\rm right} -\Delta_0^{\rm left}=\kappa
        {\Delta^{\rm right}\over a}\left({\Delta^{\rm right}\over a} -1\right) 
	+{q^2_\beta\over K_{U(1)}}- q_\beta
	\label{1.40}
        \eeq
	Similar equation has been found in the 
	 reference \cite{oz}, but with an error
	in identifying the Lorentz scale dimension.\\
	With the identifications (\ref{qek}) one sees that the KPZ equations
	in the chiral and non--chiral case are essentially the same with
	the Lorentz field being formally absorbed in the substitution
	$\Delta_0^{\rm left}\rightarrow \Delta_0^{\rm left}+
	\Delta_{\rm Lorentz}^{\rm left}$. With this in mind, we can
	continue the general discussion of the renormalized gravitational
	scaling dimensions.

    	We have at disposal four states labelled
        by $\lambda_{\alpha_\pm}$ and $\lambda_{\beta_\pm}$.
        Therefore, we can define four different renormalized
        scaling dimensions satisfying the above KPZ equation:
	\beq
        \widehat\Delta_\pm=a\left(1-{\lambda_{\beta_\pm}\over
        \lambda_{\alpha_\pm}}
        \right)
        \label{1.6}
        \eeq
        \beq
        \Delta_\pm=a\left(1-{\lambda_{\beta_\mp}\over
        \lambda_{\alpha_\pm}}
        \right)
        \label{1.6bis}
        \eeq
	
        Equation (\ref{1.6bis}) has a 
        non--trivial limit $\Delta_0\rightarrow 0$ (i.e. $\lambda_{\beta_\pm}
        \rightarrow \lambda_{\alpha_\pm}$) which allows the
        definition of a non--trivial string susceptibility coefficient as the
        renormalized scaling dimensions of the vacuum
        $\Delta_\pm\vert_{\Delta_0=0}=\Gamma_\pm$. Accordingly, we find

        \beq
        \Gamma_+=\lambda_{\alpha_-}-\lambda_{\alpha_+}\ ,
        \label{1.05}
	\eeq
        and
        \beq
        \Gamma_-  = a{\Gamma_+\over \Gamma_+ -a}\ .
        \label{1.5}
        \eeq

	\medskip
	\begin{table}[hbt]
	\smallskip
	\begin{center}
        \begin{tabular}{|p{3cm}|p{4cm}|p{4cm}|}
        \hline\hline
        \phantom{$N=0$} &\smallskip \hfill  $\Gamma_+$\hfill\smallskip 
	&\smallskip \hfill  $\Gamma_-$\hfill\smallskip\\
        \hline
        \smallskip \hfill $ N=0$  \hfill\smallskip
        &\smallskip \hfill $K+3$ \hfill\smallskip
	 &\smallskip \hfill $(K+3)/(K+2)$ \hfill\smallskip  
	 \\
        \hline
        \smallskip \hfill $N=1$  \hfill\smallskip
        & \smallskip\hfill $K+2$
        \hfill\smallskip    
	&\smallskip \hfill $(K+2)/(2K+3)$ \hfill\smallskip\\
        \hline
        \smallskip \hfill $N=2$ \hfill\smallskip
        &\smallskip \hfill $K+1$ \hfill\smallskip   
	&\smallskip \hfill $0$ \hfill\smallskip\\
        \hline\hline
 	\end{tabular}
	\caption{We display the values of the string susceptibilities 
	coefficients as described by   (\protect\ref{1.05}),
	  (\protect\ref{1.5}).  }
	\end{center}
	\label{tableg}
	\end{table}	
	\medskip
	In order to connect our result with the spectrum of physical states,
        we assume that any physical state $|\lambda \rangle $ ,
        characterized by the appropriate $SL(2,R)$ scaling dimension $\lambda$,
        is created out of a vacuum $|\lambda_\alpha \rangle $ by the action
        of some operator $O$.
        The scaling dimension of this operator $O$ is given by
        $\delta=\lambda-\lambda_\alpha$. Therefore, we can define four possible
        scaling operators $\widehat O_\pm$, $O_\pm$ with appropriate
        dimensions
	\beqa
        \widehat \delta_\pm&=&\lambda_{\beta_\pm}-\lambda_{\alpha_\pm}\ ,
	\label{1.8}\\
         \delta_\pm&=&\lambda_{\beta_\pm}-\lambda_{\alpha_\mp}\label{1.9}\\
	\eeqa
	From the general relations (\ref{1.2}) one can further prove
	\beqa
	\delta_+&=&-\delta_-\\
	\widehat \delta_+&=&-\widehat \delta_-\ ,
        \eeqa
        then,  the following relations are established
        \beqa
        \Delta_\pm &=&-{\lambda_{\alpha_+}\over \lambda_{\alpha_\pm}}
	\delta_\pm\\
	\label{scala1}
        \widehat \Delta_\pm &=&{\lambda_{\alpha_+}\over \lambda_{\alpha_\pm}}
        \widehat \delta_\pm\ .
        \label{scala2}
	\eeqa
	The string susceptibility results to be
        \beqa
        \Gamma_\pm={\lambda_{\alpha_+}\over \lambda_{\alpha_\pm}}
        \left(\widehat \delta_\pm -\delta_\pm\right)
        \label{boh}
	\eeqa
        In order to display the action of various operators,
        whose scaling dimensions are given by the above expressions,
        let us introduce the following diagram:
	\beqa
	&&\vert\lambda_{\beta_+}\rangle\qquad 
	\buildrel \displaystyle{O_+\widehat O_+}\over 
	{\longleftarrow\!\!\!-\!\!\!-\!\!\!-\!\!\!-\!\!\!-\!\!\!-}
	\quad
	\quad\quad\!\!\!\vert\lambda_{\beta_-}\rangle\nonumber\\
	\widehat O_+=\widehat O_{\Delta_+} &&\Bigg{\uparrow}
	\phantom{\vert\lambda_{\beta_+}\rangle\qquad\longleftarrow
	\qquad\qquad\quad}
	\Bigg{\uparrow}\widehat O_-=\widehat O_+^{-1}
	\nonumber\\
	&&\vert\lambda_{\alpha_+}\rangle \qquad
	\buildrel \displaystyle{O_{\Gamma_+}=\widehat O_+  O_+^{-1}}\over
	{-\!\!\!-\!\!\!-\!\!\!-\!\!\!-\!\!\!-\!\!\!\longrightarrow} 
	\>\,\,\vert\lambda_{\alpha_-}\rangle
	\nonumber
	\label{schema}
	\eeqa
	Now, it is  possible to interpret the {\it plethora} of definitions
	given by the earlier formulae. First of all, the existence
	of two different $SL(2,R)$ vacua divides physical states into a
	``$+$'' and a ``$-$'' sectors. The scale operators that induce
         transition in the each sector separately are the hatted ones .
	The transitions between different sectors is achieved by the
        un--hatted operators along the diagonals of the
	diagram, or by a combination
        of hatted and un--hatted operators along the sides. In particular
        the string susceptibility $\Gamma_+$ is represented in the above
	scheme as the scale dimension of an operator that connects the 
	different vacua. In other words, the existence of non--degenerate
	 $SL(2,R)$ vacua implies non--trivial string susceptibility
	and, therefore, phase transitions in the appropriate theory
        of gravity. On the contrary, the corresponding hatted quantity
	is trivially zero since acting
        in the same  sector. 
	In view of the additive character of the scaling dimension,
	they are truly given by the definitions (\ref{1.8}), (\ref{1.9}).
	The ``scaling'' dimensions $\Delta$'s, satisfying KPZ equation,
	are equal to true scaling dimensions $\delta$'s only for 
	``$+$''quantities (hatted and un--hatted) as can be seen in 
	(\ref{scala1}), (\ref{scala2}).
	On the other hand, $\Delta_-$, $\widehat\Delta_-$ and, as a 
	consequence, $\Gamma_-$,  cannot be interpreted in terms of the 
	simple scheme given above because, as can be seen from (\ref{scala1}),
	(\ref{scala2}), (\ref{boh}), they are multiplicatively
	related to $\delta$'s and $\widehat \delta$'s, while the property
	of genuine scale dimensions is their additivity.   
	It may be interesting to compare our approach with the
	recently proposed  modified matrix models \cite{mm1},\cite{mm2}. 
	 From our vantage point, the effect of the fine tuning of the 
	interaction in the matrix models  can be interpreted as a switching
	from the transition connecting
	the vacuum, say $\vert \lambda_{\alpha_+}\rangle $, to the
	state $\vert \lambda_{\beta_+}\rangle $, to the transition from
	the same vacuum to another state $\vert \lambda_{\beta_-}\rangle $.
	In other words, the effect of additional interaction in modified
	matrix models amounts to the exchange  $\vert \lambda_{\beta_+}\rangle
	\longleftrightarrow \vert \lambda_{\beta_-}\rangle $. 

	Although we can also  reproduce the modified string 
	susceptibility coefficient given by (\ref{1.5}), we are not able 
	to give it an interpretation within
        our scheme because of the difficulty mentioned above.

	\section{Restrictions on various parameters}
	\label{restrict}
	\subsection{Non--Chiral models}
	It is known that in the conformal gauge the parameters of the model 
	should satisfy certain conditions. One of them is the requirement 
	of reality of the parameter $\alpha$ in order to correctly define
	the scale transformation with respect to the background metric.
	The expression for $\alpha$ is obtained from (\ref{5.2}) 
	and is given by
	\beq
	\alpha_\pm={-\sqrt{m Q^2}\pm \sqrt{m Q^2-8a}\over 2\sqrt{m}}
 	\label{radici}
	\eeq
	Requirement of reality of $\alpha$ reduces to the condition
	$m Q^2-8a\ge 0$. By using the explicit form of $Q_{conf}$ one
	finds the condition
	\beq
	0\ge c_{\rm matt.}+c^{\rm conf.}_{\rm ghost}+24a+
	m +{N\over 2}
	\label{cghost}
	\eeq
	
	\medskip
	\begin{table}[htb]
	\smallskip
	\begin{center}
        \begin{tabular}{|p{3cm}|p{3cm}|}
        \hline\hline
        \phantom{$N=0$} &\smallskip \hfill  $d$\hfill\smallskip \\
        \hline
        \smallskip \hfill $ N=0 $ \hfill\smallskip
        &\smallskip \hfill $\le 1$
        \hfill\smallskip   \\
        \hline
        \smallskip \hfill $N=1$  \hfill\smallskip
        & \smallskip\hfill $\le 1$
        \hfill\smallskip    \\
        \hline
        \smallskip \hfill $ N=2$ \hfill\smallskip
        &\smallskip \hfill $\le 1$ \hfill\smallskip   \\
 	\hline\hline
	\end{tabular}
	\caption{We display the restriction on the number of scalar fields
	$d$ (``~dimension~'' of the appropriate model). It shows the well known
	$d\le 1$ barrier, characteristic of the DDK approach. Warning: $d$
	is not always identical to $c_{matt}$, as explained earlier.}
	\end{center}
	\label{table5}
	\end{table}	
	\medskip
	For transparency we give the restrictions
	on the number of matter fields, in various non--chiral
	supersymmetric models, which follow from the above condition
	in Table 5.
	Secondly, one can show, by an appropriate rescaling, that 
	the kinetic term of the Liouville field is multiplied by 
	$Q^2$, so that it results to be physical only if 
	$Q^2$ is positive \cite{tsey}.  This condition translates into
	the following equation
	\beq
	0\ge c_{\rm matt.}+c^{\rm conf.}_{\rm ghost}+
	 m +{N\over 2}
	\label{cmatt}
	\eeq
	\medskip
	\begin{table}[htb]
	\smallskip
	\begin{center}
        \begin{tabular}{|p{3cm}|p{3cm}|}
        \hline\hline
        \phantom{$N=0$} &\smallskip \hfill  $d$\hfill\smallskip \\
        \hline
        \smallskip \hfill  $N=0$  \hfill\smallskip
        &\smallskip \hfill $\le 25$
        \hfill\smallskip   \\
        \hline
	\smallskip \hfill $N=1$  \hfill\smallskip
        & \smallskip\hfill $\le 9$
        \hfill\smallskip    \\
        \hline
        \smallskip \hfill $ N=2$ \hfill\smallskip
        &\smallskip \hfill $\le 1$ \hfill\smallskip   \\
        \hline\hline
	\end{tabular}
	\caption{ We display the restrictions on the parameter $d$ 
	which follows from (\protect\ref{cmatt}). 
	It is clear that the condition 
	displayed in Table 5 are more stringent and 
	represent a 
	common solution for (\protect\ref{cghost}), (\protect\ref{cmatt}).}
	\end{center}
	\label{table6}
	\end{table}
	\medskip
	Explicit results satisfying the above condition are given in 
	Table 6. 
	
	The condition (\ref{cghost}) is more stringent  and 
	gives the common solution for both (\ref{cghost}), (\ref{cmatt}).
	It is known as ``~$d=1$ barrier~'' which restricts the application
	of the DDK approach to minimal models only. 

	On the other hand, in the light cone gauge the relevant parameter  is
	 the renormalized Kac--Moody central charge which must be real. 
	Equations (\ref{5.8}) and (\ref{kappaeq}) relate this central charge 
	to $\alpha^2$. However, real $\alpha^2$  can be obtained
	by two possible choices for $\alpha$: \\
	i) $\alpha$ real. This condition has been analyzed above, and
	the corresponding induced gravity models can be described in terms 
	of minimal models;\\
	ii)  $\alpha$ {\it imaginary}. Then, the Liouville field becomes 
	ghostlike in the conformal gauge \cite{distl}, 
	and its physical meaning turns out to be
	unclear. On the contrary, this condition  is still allowed in the 
	light--cone gauge, as discussed above. In this case $Q^2< 0$, which
	leads to the condition   

	\beq
	0\le c_{\rm matt.}+c^{\rm conf.}_{\rm ghost}+
	m+{N\over 2}
	\label{cc}
	\eeq
		\medskip
	\begin{table}[htb]
	\smallskip
	\begin{center}
	\begin{tabular}{|p{3cm}|p{3cm}|}
        \hline\hline
        \phantom{$N=0$} &\smallskip \hfill  $d$\hfill\smallskip \\
        \hline
        \smallskip \hfill  $N=0$  \hfill\smallskip
        &\smallskip \hfill $\ge 25$
        \hfill\smallskip   \\
        \hline
	\smallskip \hfill $N=1$  \hfill\smallskip
        & \smallskip\hfill $\ge 9$
        \hfill\smallskip    \\
        \hline
        \smallskip \hfill $N=2$ \hfill\smallskip
        &\smallskip \hfill $\ge 1$ \hfill\smallskip   \\
        \hline\hline
        \end{tabular}
	\caption{We display the restrictions on the parameter $d$ which follows
	from (\protect\ref{cc}). 
	We can see that in $N=2$ case there is no intermediate
	region, showing the absence of complex values of the Kac--Moody 
	central charge, as argued earlier.}
	\end{center}
	\label{table7}
	\end{table}
	\medskip
	As before, the range of explicit values is given in 
	Table 7 .
	Equation (\ref{cc}) enables one to access the region of 
	large $c_{\rm matt.}$, 	where the stringy character of induced 
	gravity is relevant
	and the inverse Kac--Moody central charge is small enough
  	to allow reliable perturbative calculations
	in the light--cone gauge. \\

	\subsection{Specific features of the chiral models}

	At this point we would like to apply the above general formulae
	to  chiral (super) gravity models. 
	The underlying motivation is the presence of the  
	Lorentz degree as an additional, dynamical, component of the
	gravitational field arising from 
	the breaking of the $U(1)$ Lorentz symmetry, and as such
	these models are different from previously analyzed non chiral
	models. In particular, 
	there is a free parameter in this theory as a result of the
	impossibility to remove the Lorentz anomaly. It is hidden 
	within $Q_{\rm Lorentz}$ as defined by (\ref{dueunoquattro}).
	This formula can be rewritten in terms of the parameters of the
	asymmetric action (\ref{tre}), with the help of relations 
	(\ref{quattro}), as \footnote{For consistency, we mention the relation
	between $Q_{Lorentz}$ as defined in (\ref{ql}) and in 
	(\ref{dueunoquattro}) 
	$3m Q_{Lorentz}^2\rightarrow 48\pi Q_{Lorentz}^2$. With this rescaling
	one gets rid of factors $\pi$ following from (\ref{quattro}).        }
	\beq
	3m Q_{\rm Lorentz}^2={\hat b\over 4}+{\Delta {\cal N}\over 2}+
	{\Delta {\cal N}^2\over 4\hat b}\ .
	\label{ql}
	\eeq
	As we have already clarified in the previous discussion, the 
	contribution of the Lorentz field will be
	contained  within $c_{\rm matt.}$ which is now  given by
	\beq
	c_{\rm matt.}^{chir.}= {{\cal N}-\Delta{\cal N}\over 2} +
	m\left(1+3Q_{\rm Lorentz}^2\right) +{N\over 2} 
	\eeq

	${\cal N}$ and $\Delta{\cal N}$ will contain the contributions of the
	(super) matter fields, and $N/2$ is the contribution of the 
	super--partners of the Lorentz degree of freedom.

	With the above definitions, (\ref{cghost}) reduces to
	a quadratic equation in the parameter $\hat b$.The roots of that 
	equation  are given by
	\beqa 
	\hat b_\pm = &-&\left[\left({{\cal N}+\Delta {\cal N}\over 2}
        +c_{\rm ghost}^{\rm conf}+24a+N+2m\right)^{1/2}\pm\right.\nonumber\\
        &&\left.\left({{\cal N}-\Delta {\cal N}\over 2} +
        c_{\rm ghost}^{\rm conf}+24a+N+2m\right)^{1/2}\,\right]^2
	\label{sol2}
        \eeqa

	In the same way (\ref{cmatt}) has roots  
	\beqa
	\hat b_\pm = &-&\left[\left({{\cal N}+\Delta {\cal N}\over 2}
        +c_{\rm ghost}^{\rm conf}+N +2m\right)^{1/2}\pm\right.\nonumber\\
        &&\pm\left.\left({{\cal N}-\Delta {\cal N}\over 2} +
        c_{\rm ghost}^{\rm conf}+N+2m\right)^{1/2}\,\right]^2
	\label{sol1}
	\eeqa
	 One can see that the equations (\ref{cc}), (\ref{cghost}),
	(\ref{cmatt}), can be always satisfied
	for any ${\cal N}$, $\Delta {\cal N}$, due to the arbitrariness 
	of $\hat b$.
	For the case $N=0$, ${\cal N}$ and $\Delta{\cal N}$ are given by
	(\ref{quattro}). It can be easily generalized to contain also
	supersymmetric matter contributions in the following way 
	\beq
	{\cal N}=md + {m\over 2}\left(n_+ + n_-\right)
	\eeq
	and 
	\beq
	\Delta {\cal N}={m\over 2}\left(n_- - n_+\right)
	\eeq
	where, $d$ represents the contribution of the scalar fields, $n_+$
	is the contribution of the left chirality fermions, and $n_-$ is
	the contribution of the right chirality fermions. Supersymmetry forces
	$n_+=d$.In order to have
	genuine chiral models, we assume that the number of  heterotic 
	(right chirality) fermions
	is different from the one given in the heterotic string models.
	This is necessary in order to have a Lorentz anomaly. So to speak,
	we are considering {\it off--critical} dimensions heterotic string
	models described as chiral gravity models.
	In order to be coherent with the notation in the non--supersymmetric 
	case (\ref{quattro}) we assumed that, in that case, fermions are
	Majorana fermions, while superfield contains Weyl--Majorana fermions.
	This is why we have $n_\pm\rightarrow 2n_\pm$ in the formula 
	(\ref{quattro})

	Explicit calculations show  that the
	equation (\ref{sol2}) depends only on the matter contribution.
	This leads to a very simple expression for the conformal gauge 
	ghost contribution  
	\beq
	c_{\rm ghost}^{\rm conf}=-24a-2m -N
	\label{ghost}
	\eeq
	With the help of the equation (\ref{cvari}), one can find an equally
	simple expression for the ghost contribution in the light--cone
	gauge
	\beq
	c_{\rm ghost}^{\rm lc}=-\left(\sqrt{6C_v} +\sqrt{dim(G)}\right)^2-
	m-{N\over 2}
	\eeq
	These equations give an amazingly simple way to calculate
	ghost contribution with no need to study ghost superfield
	structure model by model. All one needs to know is the 
	structure of the residual (super)$SL(2,R)$ symmetry of the light--cone
	gauge. With the help of (\ref{ghost}) equations (\ref{sol1}),
	(\ref{sol2}) are written as
	\beqa
	\hat b_\pm = &-&\left[\left({{\cal N}+\Delta {\cal N}\over 2}
        \right)^{1/2}\pm \left({{\cal N}-\Delta {\cal N}\over 2} 
        \right)^{1/2}\,\right]^2\nonumber\\
	\hat b_\pm = &-&\left[\left({{\cal N}+\Delta {\cal N}\over 2}
        -24a\right)^{1/2}\pm \left({{\cal N}-\Delta {\cal N}\over 2} -24a
        \right)^{1/2}\,\right]^2
	\label{sol11}
	\eeqa
	The equation (\ref{sol11}) corresponds to zeroes of the
	quadratic equation in $\hat b$ that should satisfy the requirement 
	of real $\alpha$, as in (\ref{cghost}), for chiral models.  
	First, we realize that there are real zeros for any choice of matter
	fields and the inequality (\ref{cghost}) can be
	fulfilled. But, $\hat b$ is {\it always negative.} 
	For the equation (\ref{sol2}) the previous analysis goes through
	unchanged and the conclusion is that the inequalities (\ref{cmatt})
	(\ref{cc}) can be satisfied as well. However, in this case 
	$\hat b$ can be either positive or negative depending on the
	choice of matter fields. Therefore, there is a common solution
	for the inequalities (\ref{cmatt}), (\ref{cghost}) with 
	{\it $\hat b<0$}, while inequality (\ref{cc}) can be satisfied
	with both  $\hat b>0$,  $\hat b<0$. 
	The meaning of these choices will be discussed 
	in the conclusions. Nevertheless, renormalized  Kac--Moody central 
	charge is always real in chiral models contrary to the non--chiral 
	situation.
	
	\section{Discussion of the results}

In this paper we have concentrated on the description of
the parameters characterizing the two--dimensional 
light--cone structure of induced gravity, in order to take advantage of the
underlying $SL(2,R)$ residual symmetry. We have given a general description
of the equations containing all the important information about the
theory. In this way, we were able to summarize all the chiral and non chiral
models, as well as their supersymmetric versions, within two basic formulae 
written in terms of two fundamental parameters expressible through
the characteristic quantities of the residual symmetry group. The 
motivation for such a general approach is to have a compact and simple
 expression containing all the relevant information about induced gravity
models. At every step we have however stressed and discussed the basic
difference among chiral and non--chiral versions of the model. The motivation
was to investigate alternative ways of avoiding the region of
complex values of the renormalized Kac--Moody central charge.\\
We have found that the chiral models, due to the presence of the Lorentz 
anomaly, offer a new possibility with respect to the non chiral models.
This is due to the presence of an arbitrary parameter,
which stems from a chiral theory because no regularization, i.e. fixing of the
parameter, can restore the classical symmetry \cite{anais2}. 
So far, the treatment of induced gravity models was mainly restricted to a
particular gauge choice. Comparison between the two gauges has been given
in more detail in \cite{tsey}, as far as we know. Therefore, we considered
it useful to establish detailed connection among the parameters characterizing
the two gauges. This could be useful when considering the strong gravity
region in the conformal gauge.  
Once we have established connection between the parameters of the
light--cone gauge and the corresponding parameters in the conformal gauge, 
we see that the reality of the string susceptibility in conformal gauge
is equivalent to 
the reality of the renormalized  Kac--Moody central charge
in light--cone gauge. The presence of an
arbitrary parameter in chiral models enables us to get rid of
complex values of Kac--Moody central charge for any choice of matter fields.
On the contrary, in non--chiral models only $N=2$ supergravity model
offers such a possibility due to the particular structure of the residual
$SL(2,R)$ symmetry, i.e. $dim(G)=0$. Furthermore, we have analyzed restrictions
on various parameters in both gauges, and found that the condition imposed
in the conformal gauge can be satisfied with the choice of parameter
$\hat b<0$ only. This would be the chiral analog of the $d=1$ barrier
for the non--chiral models. We have also found that the stringy region
(large $d$) of the chiral induced gravity can be reached by the choice 
of the parameter $\hat b>0$. This would be the analog of the situation
in the non--chiral models, where DDK conformal gauge approach is limited
to the minimal models only, while the light--cone approach enables 
to investigate the large $d$ region as well, since the Kac--Moody central
charge is real there too.   
This would be the end of the parallel between chiral
and non--chiral models  {\it if} there were no further, independent,
restrictions on the parameter  $\widehat b$. However, if one looks at the
structure of the Lagrangian (\ref{local}) written in terms of the
dynamical components in the light--cone gauge, one can see that the parameter
 $\widehat b$ figures as a coupling constant of the kinetic term of the 
Lorentz field. Therefore, apart from the requirement of the reality of 
various physical parameters, one has additionally to require the correct 
sign of the kinetic term of the Lorentz field to make it physical.
Such a requirement puts an {\it additional} constraint on the free 
parameter $\widehat b$. In fact, in order for the Lorentz field to be 
physical one has to require $\widehat b>0$.
Comparing this constraint with the independent one following from equation 
(\ref{sol11}), which guarantees the reality of Kac-Moody central charge,  
we see that a consistent chiral gravity model,  
can only be achieved for the particular chiralities assignment 
\beq
{{\cal N}\pm \Delta{\cal N}\over 2}\le 24a
\label{final}
\eeq
For sake of transparency, we present the following table of explicit 
results for various chiral models. 
         \medskip
	\begin{table}[htb]
	\smallskip
	\begin{center}
        \begin{tabular}{|p{3cm}|p{3cm}|p{3cm}|}
        \hline\hline
        \phantom{$N=0$} &\smallskip \hfill  $n_+ =d$\hfill\smallskip
        &\smallskip \hfill  $n_-$\hfill\smallskip\\
        \hline
        \smallskip \hfill  $N=0$  \hfill\smallskip
        &\smallskip \hfill $\le 24$\hfill\smallskip
        &\smallskip \hfill $\le 24-n_+$\hfill\smallskip  \\
        \hline
        \smallskip \hfill $N=1$ \hfill\smallskip 
	&\smallskip \hfill $\le 12 $\hfill\smallskip
        & \smallskip\hfill $\le 12-n_+$\hfill\smallskip  \\
        \hline
	\smallskip \hfill $N=2$ \hfill\smallskip
        &\smallskip \hfill $\le 0$ \hfill\smallskip
         &\smallskip \hfill $\le -n_+$ \hfill\smallskip   \\
        \hline\hline
        \end{tabular}
	\caption{We list regions of allowed values of chiralities
	 leading to the consistent (with physical Lorentz degree of
	freedom) chiral models except in the case $N=2$.}
	\end{center}
	\label{table8}
	\end{table}
        \medskip
The meaning of this result is the following: although
it is possible to get rid of complex values of the renormalized
Kac--Moody central charge without restrictions on the
number of chiralities, further requirement of positive kinetic energy 
for the physical 
degrees of freedom, nevertheless restricts the number of chiralities, as it is 
described above. In this way, it is  possible to have sensible chiral gravity
models in spite of their anomalous character and, at the same time,
to avoid the unpleasant
region of complex values of renormalized Kac--Moody central charge.
With respect to the previous discussion, regarding parameter $\hat b$,
we can now say the following: additional requirement of a physical character
of the Lorentz field points to the applicability of the chiral models
to the region of large $d$ where its stringy character is relevant,
and where reasonable perturbative calculations can be performed.
$N=2$ case has a particular property: the appropriate non--chiral
model gives real Kac--Moody central charge for any value of $d$, while
the chiral version is not consistent, i.e. in this case Lorentz field
is ghostlike.

We have also found a useful generalized way of extracting essential
information about various aspects of induced gravity models.
Previously, one had to control the details of each model in terms of its
(super)field structure, (super)current, (super)ghosts, etc.  We have been 
able to give simple formulae, not previously present in the literature,
 where the above mentioned work is 
reduced to a minimum, and all one needs to know is the structure of the 
group of residual symmetry in the light--cone gauge in order to reproduce
all known results. Furthermore, we have established relations among
various parameters in the light--cone and conformal gauges, which enable us
to reproduce also the known conformal gauge results. As a byproduct of the
above results, we have given a simple derivation of the KPZ equation
in both chiral and non--chiral situations,
which led to the expression of the renormalized gravitational scaling 
dimension, and therefore of the critical exponents, in terms of the $SL(2,R)$
scale weights of the vacuum state. In this picture the string susceptibility
is described, at least in the light--cone gauge language,
 as the scale weight of an operator interpolating between
the $SL(2,R)$ vacua states, whose multiplicity gives origin to the phase 
transitions in the induced gravity models. In this framework we have 
reproduced recent results of the modified matrix models.


\begin{thebibliography}{99}


\bibitem{Pol1} A.M.Polyakov
Phys.\ Lett.\ {\bf 103B}, (1981), 207
\bibitem{gn} J-L.Gervais, A.Neveu
Nucl.\ Phys.\ {bf B238}, (1984), 396
\bibitem{tsey} A.A.Tseytlin
Int.\ J.\ Mod.\ Phys.\ Lett.\ {\bf A5}, (1990), 1833
\bibitem{curt} T.Curtright, C.Thorne
Phys.\ Rev.\ Lett.\ {\bf 48}, (1982), 309;\\
J.Gervais, A.Neveu
Nucl.\ Phys.\ {bf B199}, (1982), 59
\bibitem{ger95} J-L.Gervais, J.Schnittger
Phys.\ Lett.\ {\bf 315B}, (1993), 258\\
J-L.Gervais, J-F.Roussel
Nucl.\ Phys.\ {bf B426}, (1994), 140
\bibitem{otto} H.Dorn, H-J.Otto
in Proceed.\ XXVIII\ In.\ Symp.\ on\ the\ Theory\ of\ Elementary\ Particles\
Wendisch-Rietz,\ August\ 1994 
\bibitem{Pol3}  A.M.Polyakov
Mod.\ Phys.\ Lett.\ {\bf A2}, (1987), 893
\bibitem{Pol4} V.G.Knizhnik, A.M.Polyakov, A.B.Zamolodchikov
 Mod.\ Phys.\ Lett.\  {\bf A3}, (1988), 819
\bibitem{Pol5} A.M.Polyakov, A.B.Zamolodchikov
Mod.\ Phys.\ Lett.\  {\bf A3}, (1988), 1213\\
Z.Yang
\ Phys.\ Lett.\  {\bf B245}, (1990), 92\\
Riu-Ming Xu
\ Phys.\ Lett.\  {\bf B247}, (1990), 295
\bibitem{jr} R.Jackiw, R.Rajaraman
Phys.\ Rev.\ Lett.\ {\bf 54}, (1985), 1219
\bibitem{anais1} A.Smailagic
Phys.\ Lett.\ {\bf B195}, (1987), 213
\bibitem{distl} J.Distler, H.Kawai
Nucl.\ Phys.\ {\bf B321}, (1989), 509;\\
F.David
Mod.\ Phys.\ Lett.\ { \bf A3}, (1988), 1651
\bibitem{cqg}S.J.Gates jr., M.T.Grisaru, L.Mezincescu, P.K.Townsend
Nucl.\ Phys.\ {\bf B286}, (1986), 1;\\
 A.Smailagic, E.Spallucci
Class.\ and\ Quant.\ Grav.\ {\bf 10}, (1993), 451;
\bibitem{anais2} A.Smailagic, E.Spallucci
Phys.\ Lett.\ {\bf B284}, (1992), 17
\bibitem{leut} H.Leutwyler
\ Phys.\ Lett.\ {\bf B153}, (1985), 65
\bibitem{reut} A.H. Chamseddine, M. Reuter
Nucl.\ Phys.\ {\bf B317}, (1989), 757
\bibitem{smsp} A.Smailagic, E.Spallucci
\ Phys.\ Lett.\ {\bf B317}, (1993), 526
\bibitem{oz2} K.A.Meissner, J.J.Pawelczyk
 Mod.\ Phys.\ Lett.\  {\bf A5}, (1990), 763\\
A.Smailagic,
Mod.\ Phys.\ Lett.\  {\bf A5}, (1990), 125\\
G.W.Delius, M.T.Grisaru, P.van Nieuwenhuizen
Nucl.\ Phys.\ {\bf B389}, (1993), 25 
\bibitem{Pol6} A.M.Polyakov, 
in Field, Strings and Critical Phenomena,\  (1989), ed.
E.Brezin, J.Zinn-Justin, Les Houches
\bibitem{fre} V.G.Knizhnik, A.B.Zamolodchikov
Nucl.\ Phys.\ {\bf B247}, (1984), 83\\
P.Fre, F.Gliozzi
\ Nucl.\ Phys.\ {\bf B326}, (1989), 411
\bibitem{conf} A.A. Belavin, A.M.Polyakov, A.B.Zamolodchikov
\ Nucl.\ Phys.\ {\bf B241}, (1984), 333
\bibitem{ung} Z.Horvath, L.Palla, P.Vecsernyes
Int.\ J.\ Mod.\ Phys.\ {\bf A4}, (1989), 5261;
\bibitem{11} T.Kuramoto
Nucl.\ Phys.\ {\bf B346}, (1990), 527
\bibitem{dock} E.D' Hocker, R.Jackiw
Phys.\ Rev.\ {\bf D26}, (1982), 3517
\bibitem{li} J.Kim, T.Lee
Phys.\ Rev.\ {\bf D42}, (1990), 26
\bibitem{mp} 
R.C.Myers, V.Periwal,
Nucl.\ Phys.\ {\bf B397}, (1993), 239
\bibitem{ddk2} J.Distler, Z.Hlousek, H.Kawai
Int.\ J.\ Mod.\ Phys.\ {\bf A5}, (1990), 391
\bibitem{oz} Y.Oz, J.J.Pawelczyk, S.Yankielowicz
 Nucl.\ Phys.\ Lett.\  {\bf B363}, (1991), 555
\bibitem{mm1} I.R.Klebanov
Phys.\ Rev.\ {\bf D51}, (1995), 1836
\bibitem{mm2}I.R.Klebanov, A. Hashimoto
Nucl.\ Phys.\ {\bf B434}, (1995), 264
\end{thebibliography}
\end{document}